\documentclass[11pt,twoside]{article}

\usepackage{asp2006}
\usepackage{graphicx}
\usepackage{epsfig}
\usepackage{lscape}

\begin{document}
\title{Subluminous O Stars -- Origin and Evolutionary Links}
\author{Uli Heber}
\affil{Dr. Remeis-Sternwarte, Universit\"at Erlangen-N\"urnberg, 
Sternwartstr. 7, 96049 Bamberg, Germany}

\begin{abstract}
Hot subluminous stars can be roughly divided into B- and O-types. Unlike the
latter many sdBs are found in close binaries, indicating that binary evolution
plays a vital role.  
Recent NLTE spectral analyses revealed that
an evolutionary link between sdB stars and sdO stars is plausible only 
for the \emph{helium-deficient} sdO stars, i.e\@. 
they are the likely successors to 
sdB stars.  The atmospheric properties
of \emph{helium-enriched} sdO stars can be
explained by the late hot flasher as well as by the white-dwarf 
merger scenarios, although both models do not match 
the observed properties of \emph{helium-enriched} sdO stars in detail. 
The white dwarf merger scenario is favoured because it naturally explains the 
scarcity of
close binaries amongst \emph{helium-enriched} sdO stars. A hyper-velocity sdO star
moving so fast that it is unbound to the Galaxy has probably been ejected by the
super-massive black hole in the Galactic centre.  
\end{abstract}

\section{Hot Subluminous Stars}

Hot subluminous stars are an important population of faint blue stars at high
Galactic latitudes closely related to the horizontal branch.
A proper spectral classification of hot subluminous stars is rendered
difficult by the diversity of the helium line spectra. They can be
grouped roughly into 
the cooler sdB stars, whose spectra typically display no or only weak helium
lines, and the hotter sdO stars, which have a higher helium abundance on
average and can even be dominated by helium.
The former have recently been studied extensively because they are
common enough to account for the UV excess observed in early-type
galaxies. Pulsating sdB stars are important tools for asteroseismology
\citep{char04}, and sdB stars in close binaries may qualify as supernova Ia
progenitors \citep{maxt00,geier07}. 

Subluminous B stars have been identified as extreme horizontal branch (EHB)
stars
\citep{heber86}; i.e., they are core helium-burning stars with hydrogen envelopes
that are 
too thin to sustain hydrogen burning (unlike normal HB stars).
Therefore they evolve directly to the white-dwarf cooling sequence by avoiding 
the asymptotic giant branch (AGB).
While the sdB stars spectroscopically form a homogeneous class, a large variety 
of spectra is observed among sdO stars \citep{heb92,heb06}. 
Most subluminous B stars are helium poor, whereas only a relatively
small
fraction of sdO stars are.

Ever since the pioneering work by \citet{green74},
the helium-rich sdO stars were believed to be linked to
the evolution of the hydrogen-rich subluminous B stars. 
Any evolutionary link between subluminous B and O stars, however, is
difficult to explain since the physical processes
driving a transformation of a hydrogen-rich star into a helium-rich one
remain obscure. The convective transformation has been explored by 
\citet{wese81}, as well as by \citet{grot85}. While the former
found helium convection even at subsolar helium abundances, which
mixes helium from deeper layers into the photosphere, the
latter concluded that a helium-driven convection zone develops only in
helium-rich atmospheres. If the latter is true, convective transformation
would not work.

The fraction of sdB stars in short period binaries (periods less than ten days) is high.
\citet{max01} found 2/3 of their sdB sample were such binaries, whereas 
a somewhat lower fraction of 40\% was found recently for 
  the sample drawn from ESO Supernova Ia Progenitor SurveY
  \citep[SPY, ][]{napi01}. Quite to the opposite, radial velocity 
  variable 
stars are rare amongst the \emph{helium-enriched} sdOs, for which 
\citet{napi04} find that a fraction of radial velocity variables to be
 4\,\% at most.
Obviously, binary evolution plays an important role in the formation of sdB 
stars and possibly also in that of the sdO stars. 

While atmospheric parameters have been determined for several hundred
sdB stars \citep[e.g. ][]{saffer94, max01, edelmann03, lis05}, only very few 
sdO stars have been analysed so far. Most of the apparently brightest sdO stars
turned out to be post-AGB stars \citep[e.g. ][]{rauch91, heber88} and shall not
be discussed here.
LTE model atmospheres are sufficient to
analyse the B-type subdwarfs \citep{napiwotzki97}, whereas NLTE is mandatory for the sdO stars
rendering the analysis more difficult. Early NLTE analyses of low resolution
spectra gave inconsistent results \citep{dreizler90,thejll94}.  
The SPY survey has 
provided high resolution spectra of 46 sdO stars and a new grid of model
atmospheres has become available \citep{stroer07}. \cite{hirsch08} have
recently used the same grid to analyse sdO spectra from the Sloan Digital Sky 
Survey (SDSS).
Hence, atmospheric parameters for about 130 sdO stars are now at hand, a
sufficiently large number to test rivalling evolutionary scenarios.
We describe the NLTE spectral analyses in the next section and compare the
results from the SPY survey to evolutionary models in section 3. In section 4 we
add the results from the SDSS survey. Before concluding we present the discovery
of a so-called hyper-velocity sdO star in section 5.

\section{NLTE Spectral Analysis of the SPY Sample}

The SPY project has obtained high resolution spectra with the 
UVES spectrograph at the ESO-VLT for over 1000 white-dwarf candidates to test
possible scenarios for type Ia supernovae by searching for double
degenerate white-dwarf binary systems close to the Chandrasekhar mass
limit. Many of the target stars of SPY came from the Hamburg ESO survey. 
SPY also observed 137 hot subluminous stars that entered the
target sample because they were previously classified as white dwarfs. 
Seventy-six of these stars are now classified as sdB/sdOB, and 58
as O-type subdwarfs.  

\cite{stroer07} determined atmospheric parameters ($T_{\mathrm{eff}}$, 
$\log g$\ and $\log y$) 
of 46 sdO stars from SPY by fitting synthetic model spectra to the observed 
ones, using a $\chi^2$-minimisation procedure \citep{napi99} to derive all 
three parameters simultaneously.
The synthetic non-LTE models were constructed using the TMAP code 
\citep{hwerner99} with a new 
temperature correction technique \citep{dreizler03} and include partially line blanketing.
It includes H and He atoms only.
The model grid ranges in temperature from 30\,000\,K to 100\,000\,K and 
from 4.8\,dex to 6.4\,dex for $\log g$\ with $\log y = -4 \ldots +3$.
No extrapolation beyond the model grid was allowed. 
For further details see \cite{stroer07}.

\begin{figure}\centering
\resizebox{0.75\hsize}{!}{\includegraphics{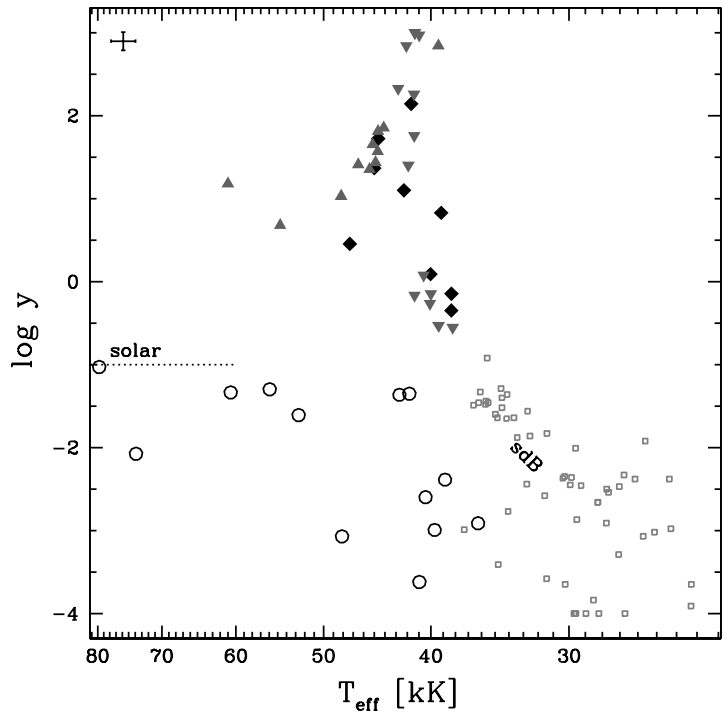}}
\caption{He/H ratio $y$ versus $T_{\mathrm{eff}}$\ for sdO 
and comparison to the sdB stars (open
squares) from the 
SPY project.
Open circles: No C and N lines; filled triangle: either N (upside down triangle)
or C lines; filled diamonds: both C and N lines visible. 
Measurement uncertainties
are given in the upper left corner \citep[from ][]{stroer07}.}
\label{results_spy_he}
\end{figure}

\begin{figure}\centering
\resizebox{0.70\hsize}{!}{\includegraphics{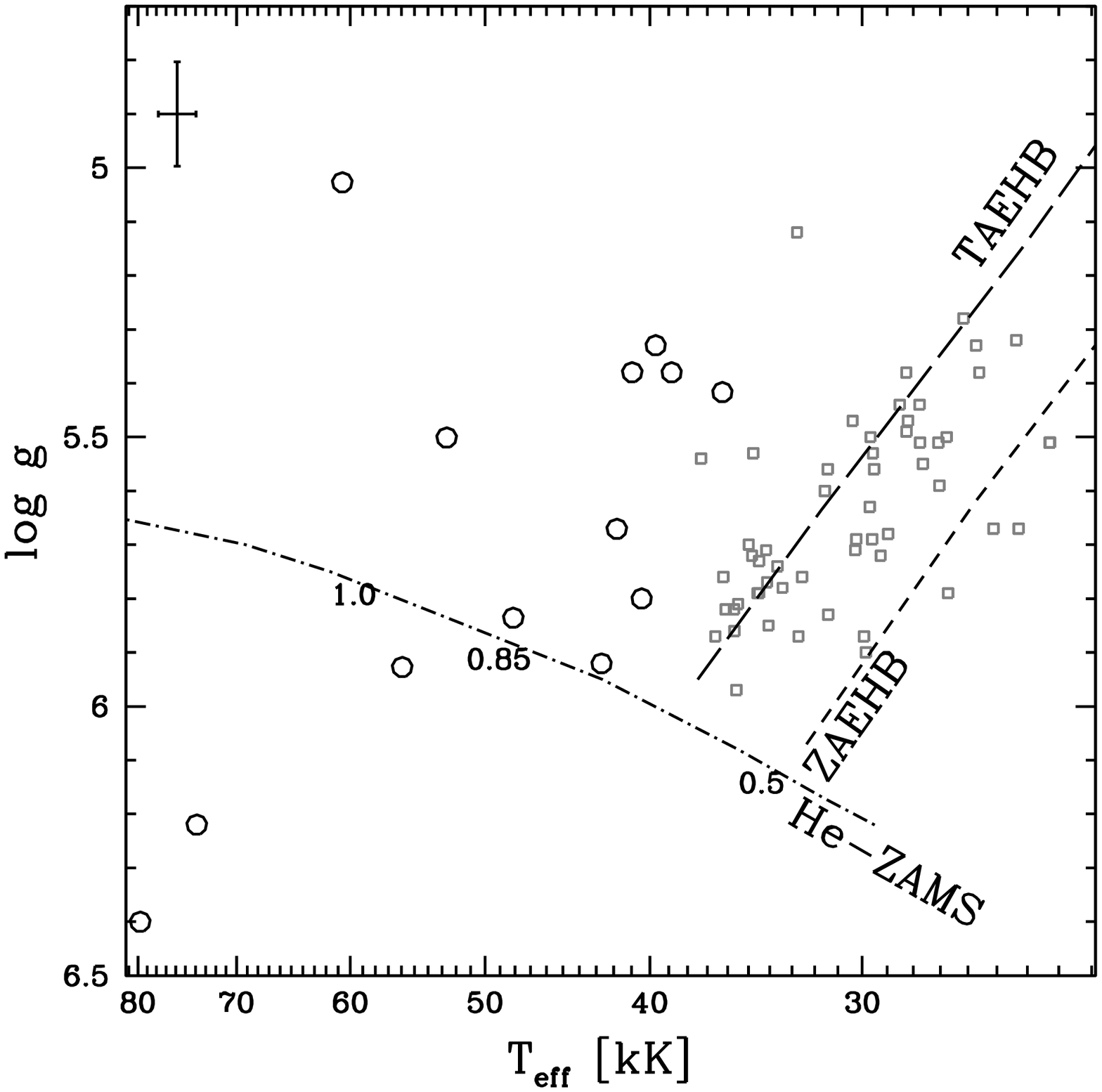}}
\resizebox{0.70\hsize}{!}{\includegraphics{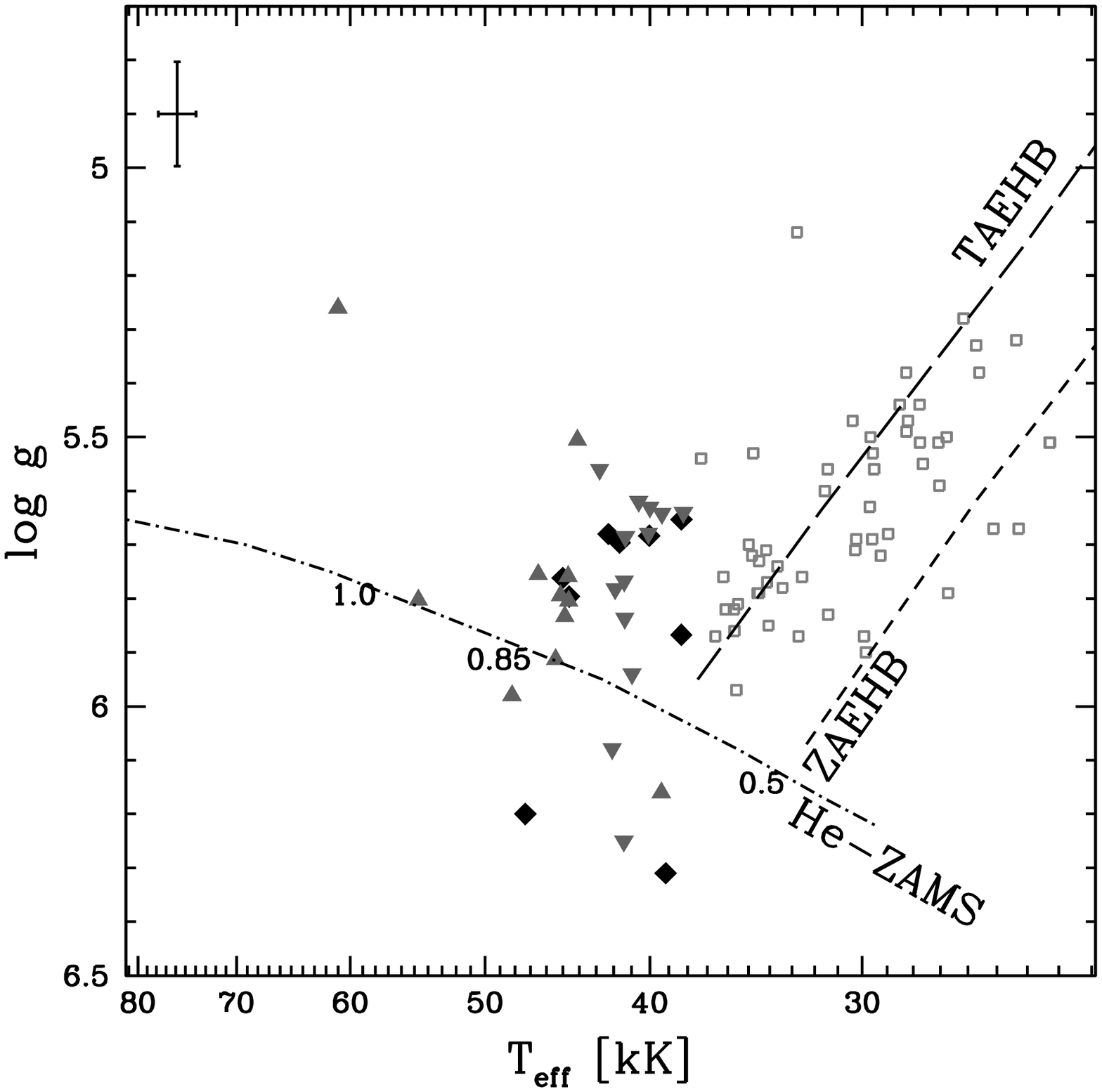}}
\caption{
Top panel: \emph{Helium-deficient} sdO stars from the SPY project: 
Distribution of $T_{\mathrm{eff}}$\ and $\log g$.
The EHB band and the helium zero-age main sequence
are also 
indicated. Notation as in Fig.~\ref{results_spy_he}.
Bottom panel: Same, but for \emph{helium-enriched} sdO stars
\citep[from ][]{stroer07}.
}
\label{results_spy_teff_logg}
\end{figure}

The results of the analysis are displayed in
Figs.~\ref{results_spy_he} and \ref{results_spy_teff_logg}.
Most strikingly a clear correlation between helium abundance and CN class 
becomes apparent.
None of the sdO stars with subsolar helium content shows carbon and/or 
nitrogen lines. 
The opposite is true for sdO stars
with supersolar helium content -- all of them show 
carbon and/or nitrogen lines.

This suggests that the sdO stars should be grouped into two classes according 
to helium content. Those with supersolar helium abundances will be
referred to as \emph{helium-enriched} sdO stars, while those 
with subsolar helium abundances will be termed
\emph{helium-deficient}
sdO stars.

While the \emph{helium-deficient} sdO stars are scattered in a wide
$T_{\mathrm{eff}}$-$\log g$-range, most \emph{helium-enriched} sdOs populate
a relatively narrow region ($T_{\mathrm{eff}}$ from $\sim$40 to $\sim$46\,kK and $\log g$
from $\sim$5.5 to $\sim$5.9).

\section{Evolutionary Scenarios: Late Hot Flashers, Common Envelope Ejection and
White Dwarf Mergers }

\subsection{Late Hot Flashers}

\begin{figure}\centering
\resizebox{0.75\hsize}{!}{\includegraphics{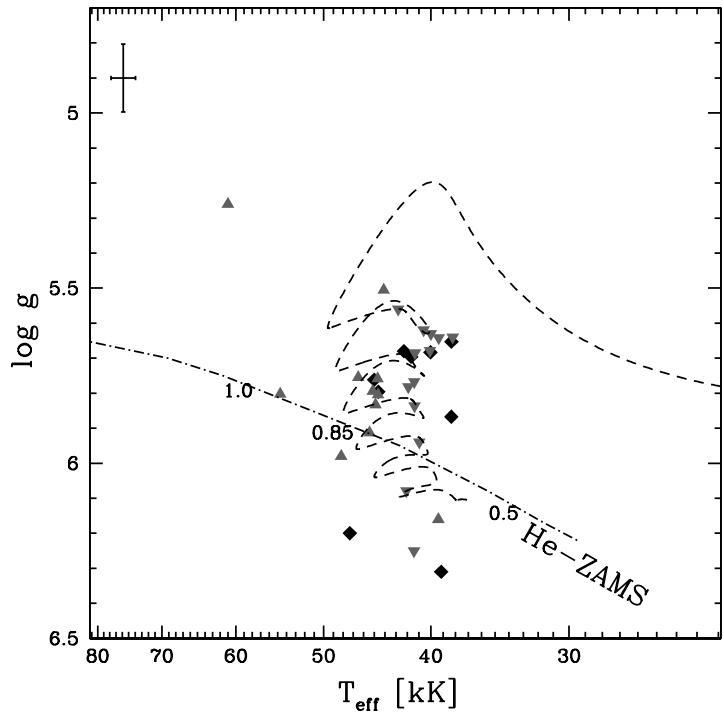}}
\caption{
\emph{Helium-enriched sdO stars}: Comparison with an evolutionary track for an 
EHB star formed by a delayed helium flash in the $T_{\mathrm{eff}}$-$\log g$
plane. The track settles onto
the helium main sequence (labelled by stellar mass in solar units). 
Symbols as in Fig.~\ref{results_spy_he} \citep[from ][]{stroer07}.}
\label{latehe}
\end{figure}

Non-standard evolutionary models were introduced to explain the formation of
sdO stars \citep[e.g.][]{swei97,brown01,moe04}. In particular, 
the {\it late hot flasher scenario} predicts that the core helium flash may
occur
when the
star has already left the red giant branch (RGB) and is approaching the 
white-dwarf cooling sequence (delayed He core flash). During the flash, He 
and C may be dredged-up to the surface. Hydrogen is mixed into deeper layers 
and  burnt. The
remnant is found to lie close to the helium main sequence, i.e\@. at the very end
of the theoretical extreme horizontal branch. In Fig.~\ref{latehe} the observed distribution of all sdB and sdO stars 
from SPY in the
$T_{\mathrm{eff}}$-$\log g$-diagram is compared to an evolutionary track settling to the EHB 
\citep{swei97}.

The final composition of the envelope is helium-dominated, 
and enriched with carbon  
\citep[or nitrogen if the hydrogen burning during the helium
flash phase burns $^{12}$C into $^{14}$N; ][]{swei97}. 
Indeed, most of our observed \emph{helium-enriched} sdO 
stars lie near the model track, suggesting that this scenario may be viable. 
However, the evolutionary time scales 
(1.95$\times 10^6$ yrs for the evolution shown in Fig.~\ref{latehe}) are much shorter than for
the helium-burning phase \citep{swei97}. 
Accordingly the stars should accumulate near the end of the track, i.e\@. near the
helium main sequence, which is not the case for our program stars. 

Although the late hot-flasher scenario can explain the helium enrichment and 
the line strengths of C and/or N lines as due to dredge up, it fails to
reproduce the distribution of the stars in the $T_{\mathrm{eff}}$-$\log g$-diagram in detail.

\begin{figure}[htb!]\centering
\resizebox{0.75\hsize}{!}{\includegraphics{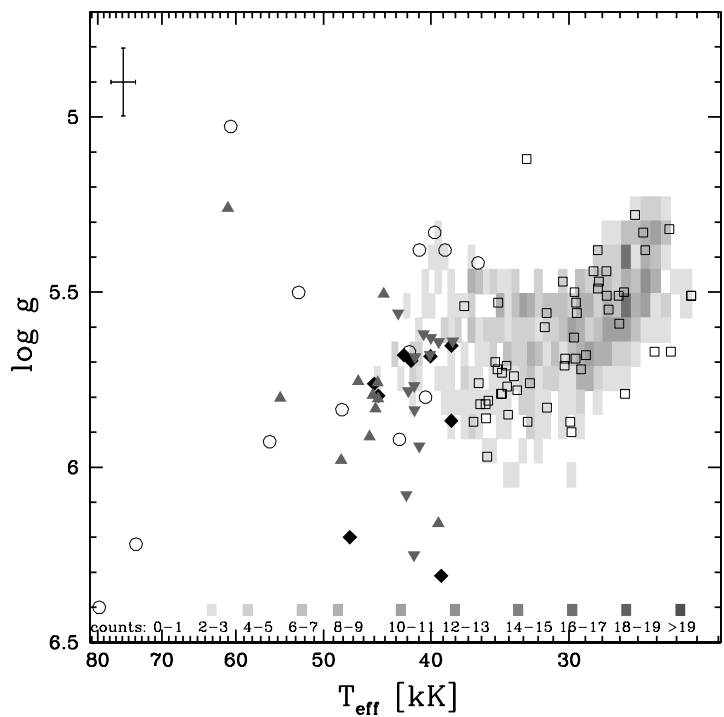}}
\caption{Comparison of the atmospheric parameters of hot subdwarfs from SPY 
to simulation set No. 10 
of \citep{han03}. 
Shaded
$T_{\mathrm{eff}}$\,-$\log g$ boxes: theoretical predictions, where a higher subdwarf density per box 
corresponds to
darker shading (grey scale shown below). 
Notation same as in Fig.~\ref{results_spy_teff_logg} \citep[from ][]{stroer07}.
}
\label{img:bin}
\end{figure}

\subsection{White Dwarf Mergers}

Evidence has accumulated that close binary evolution is important to
understand the origin of the hot subdwarf stars. 
A recent binary population synthesis study \citep{han03} 
identified three channels 
for forming sdB stars:
(i) one or two phases of common envelope evolution,
(ii) stable Roche-lobe overflow, and
(iii) the merger of two helium-core white-dwarfs.
The latter could explain the population of single stars. Short period binary WDs will lose orbital energy through gravitational waves.
With shrinking separation, the less massive object will eventually be 
disrupted and accreted onto its companion, leading to helium ignition.
\citet{saio2000} argue, that this merger product will result in a helium 
burning subdwarf showing an atmosphere enriched in CNO-processed matter.
This scenario therefore can explain these extremely \emph{helium-enriched} sdOs 
showing strong nitrogen lines in their atmospheres.
However, \citet{gour06} find that, under the assumption of total angular 
momentum conservation, He+He WD mergers do rotate faster than breakup velocity.
A mechanism that enables the star to get rid of its angular momentum still has 
to be found.

In Fig.~\ref{img:bin} the observed distribution of all sdB and sdO stars 
from SPY in the
$T_{\mathrm{eff}}$-$\log g$-diagram is compared to the simulation set No. 10 of \cite{han03},
which is characterised by a low efficiency ($\alpha_\mathrm{CE} =
\alpha_\mathrm{th} =  0.5$), low metalicity
($Z  = 0.004$), and a constant mass ratio of the progenitor binaries.
This set was chosen because it came closest to the SPY-sdB distribution \citep{lis05}.  
The simulation set is represented by
two-dimensional bins in Fig.~\ref{img:bin}.
The grey shading of the rectangular areas corresponds to the respective
number of simulated stars they contain. Higher
number densities of simulated subdwarfs correspond to darker grey
shading. 

>From the direct comparison of the $T_{\mathrm{eff}}$-$\log g$-values to the \cite{han03}
simulations, two effects become apparent. First, sdO stars significantly exceed
even the hottest
temperatures predicted.
Second, by restricting our 
analysis to stars that come close to the \cite{han03} predictions, i.e.
those that are apparently connected with the sdB sample, a 
disagreement of the observational data with the simulation set becomes
obvious: the relative amount of hot (sdO) and cool (sdB) stars differs
significantly. The binary population synthesis models have a large number of
parameters which have to be constrained by additional observations (see
Sect\@. \ref{sec:sdss}). Moreover further refinement of the binary population 
synthesis models
is required as well as an extension of parameter studies.

The hottest stars in the simulation set are mostly formed from white dwarf 
mergers. Close binaries are much rarer amongst \emph{helium-enriched} sdO 
stars than amongst the sdB stars in the SPY sample (by a factor of ten). Therefore it is tempting to
identify the \emph{helium-enriched} as formed by mergers of helium white dwarfs.

\begin{figure}[htb!]\centering
\resizebox{0.75\hsize}{!}{\includegraphics{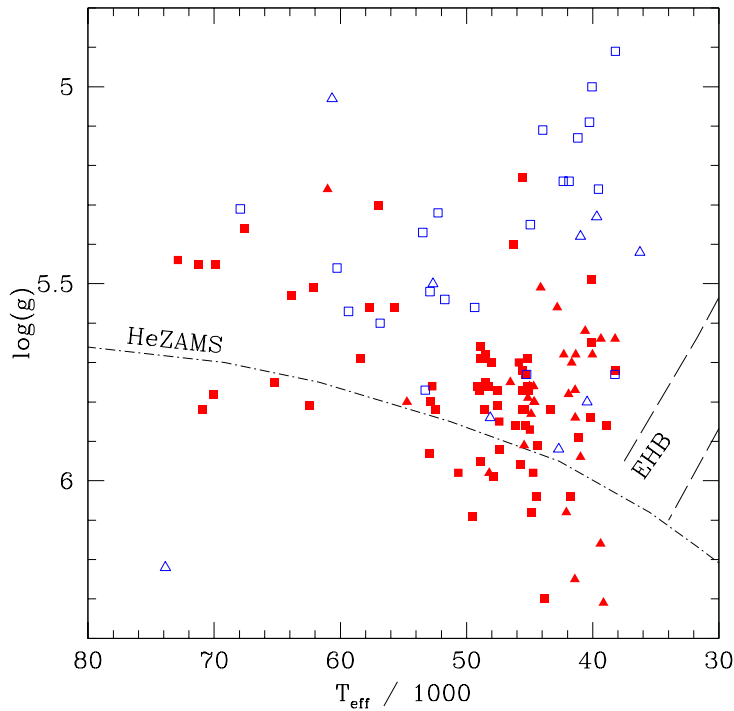}}
\caption{$T_{\mathrm{eff}}$-$\log g$-diagram: 
Filled symbols: \emph{helium-enriched} sdO; open symbols:
\emph{helium-deficient}  sdO.
Squares: SDSS, triangles: SPY \citep[from ][]{hirsch08}.}
\label{results_sdss_spy}
\end{figure}

\section{Subluminous O stars from the Sloan Digital Sky Survey}\label{sec:sdss}

\citet{hirsch08} searched the database of the 5th data release of SDSS, 
and selected all point sources within the colour box $(u-g)<0.4$ and $(g-r)<0.1$.
They classified more than 8\,000 spectra by visual inspection, mostly white 
dwarfs, about 500 sdB stars and 112 sdO stars.
After removing too noisy spectra and those with spectral features indicative of a cool 
companion, a spectral analysis of 
the remaining 87 sdO stars was performed using the same analysis technique and 
NLTE model atmosphere 
grid as described above.
The distribution of SDSS stars in the $T_{\mathrm{eff}}$-$\log g$-plane 
(see Fig.~\ref{results_sdss_spy}) resembles that of the sdOs from SPY quite
well.  

The ratio of \emph{helium-enriched} stars to \emph{helium-deficient}  ones appears to be
slightly higher in SDSS than in SPY (68:18 vs 33:13).
In addition, the SDSS sample contains more very hot stars 
($T_{\mathrm{eff}} > 60\,000$) than the SPY sample.
Both findings are probably due to selection effects, e.g. the limiting 
magnitude of the surveys, which is considerably fainter for SDSS than for 
SPY.

In summary, the increased sample consisting of the SPY and the SDSS sample
corroborates the conclusions drawn from the smaller SPY sample.  

\cite{stroer07} pointed out that the SPY sample
contains a few stars lying below the helium main sequence, all but one are 
\emph{helium-enriched}. The number was too small to draw firm conclusion. As
can be seen from Fig.~\ref{results_sdss_spy} there are also such stars in the
SDSS sample confirming the significance of the discovery (unless there is a
systematic error in the gravity determination of all of them).
The origin of stars below the helium main sequence is difficult to explain as 
such stars cannot sustain stable 
helium burning in their cores.

\section{Discovery of an Unbound Hyper-Velocity SdO star}

Amongst the sdO stars drawn from the SDSS data base,   
\cite{hirsch05} discovered a so-called hyper-velocity star,
US~708, in the Milky Way halo, with a heliocentric radial velocity of 
+708$\pm$15~$\mathrm{km\,s^{-1}}$. 

A quantitative NLTE model atmosphere analysis of optical spectra obtained 
with the KECK I telescope
shows that US~708 is a normal \emph{helium-enriched} sdO 
with $T_{\rm eff}$=44\,500~K, $\log g =5.25$. Adopting the canonical mass of
half a solar mass from evolution theory the corresponding distance is 19~kpc. 
Its Galactic
rest frame velocity is at least 757~$\mathrm{km\,s^{-1}}$, 
much higher than the local Galactic escape velocity (about
430~$\mathrm{km\,s^{-1}}$)
indicating that the star is unbound to the Galaxy.
It has been suggested by \cite{hills88} that such
hyper-velocity
stars can be formed by the tidal disruption of a binary through 
interaction with
the super-massive black hole at the Galactic centre (GC).
Numerical kinematical experiments
are carried out to reconstruct the path of US~708 from the GC. 
US~703 needs about 36~Myrs to travel from the GC to its present
position, which is shorter than its evolutionary lifetime. Hence it is plausible
that the star might have originated from the GC, which can be tested by
measuring accurate proper motions.
A HVS survey has increased the number of known HVS to ten
\citep{brown07}. However, 
US~708 remains the only bona-fide old, low mass HVS star, while all other are
probably young massive stars.

\section{Summary and Conclusion}

According to recent quantitative spectral analyses 
  the sdO stars should be grouped into two classes according 
to helium content because of a pronounced dichotomy of the carbon/nitrogen
spectra. At supersolar helium abundances (\emph{helium-enriched}) all sdO 
stars display C and/or N lines, while none are observed at  
subsolar helium abundances (\emph{helium-deficient} sdOs).
A direct evolutionary linkage of the hot sdO stars to the somewhat cooler sdB
stars is plausible only 
for the \emph{helium-deficient} sdO stars, i.e\@. 
the latter are the likely successors to 
sdB stars.

Most of the \emph{helium-enriched} sdO stars cluster in a narrow region
of the $T_{\mathrm{eff}}$-$\log g$-diagram at temperatures between 40kK and 50kK.
While diffusion is probably causing helium deficiency,
it is unlikely to account for the helium enrichment. Non-standard
evolutionary scenarios had therefore to be invoked.
The predictions from both the late hot-flasher scenario and 
the helium white-dwarf merger scenario are roughly consistent with
the observed distribution of \emph{helium-enriched} sdO stars but do not
match them in detail. 
The occurrence of both a delayed helium core flash and the merger of two
helium white
dwarfs may explain the helium enrichment.
In these cases carbon and/or nitrogen
can be dredged up to the stellar surface, which would
explain the strength of the C and/or N lines in \emph{helium-enriched} sdO 
stars. The lack of close binaries amongst the latter is consistent with a white
dwarf merger
origin.

Some high gravity \emph{helium-enriched} sdO stars lie
below the helium main sequence, which is at variance with any core helium 
burning
model. Whether the so-called He-sdB stars \citep{hahmad03} form a
 class of their own or are just the low temperature tail of the
 \emph{helium-enriched} sdO stars remains to be studied.
Detailed spectral analyses of high resolution spectra are urgently needed to 
determine C and N abundances and to derive projected rotational velocities. 
Both will yield tight constraints to test evolutionary scenarios.

The nature of the enigmatic double sdO star PG~1544+488
\citep{ahmad04} needs to be explored
further. \cite{ahmad04} argue that the mass of the secondary is unusually low, 
i.e\@. too low to sustain core-helium burning. A similar
sdO binary has been discovered by \cite{lisker04} but has not been followed up
yet.

The discovery of a hyper-velocity sdO star (US~708) travelling so fast that it is
unbound to the Galaxy came much to a surprise and the star remained the
only low mass, evolved object of its class up to now. It is generally
believed that such stars originate from the Galactic centre, because only 
super-massive black holes seem to be capable of accelerating stars to velocities
larger than the Galactic escape velocity. Because US~708 probably is a very old
star it might have been ejected from a globular cluster rather than from the GC
if a black hole of intermediate mass (a few thousand solar masses) resides in
the cluster \citep[see ][]{heber08}.     

Almost ten years after the discovery of the first multi-periodic pulsating sdB
star \citep{kilkenny97}, the first sdO star was found to pulsate \citep{woudt06}. This opens up 
a new window to study sdO stars using asteroseismological models. For subluminous
B stars this has already resulted in the measurements of masses of the stars and
their envelopes in more than a handful of cases \citep{fontaine08}. This
technique may be very promising for sdO stars as well although appropriate 
models are still at their infancy.   



\end{document}